\documentclass[11pt]{article}

\usepackage{array}
\usepackage{amsfonts,amsmath,amssymb,amsthm,mathrsfs}
\usepackage{adjustbox}
\usepackage{tikz}
\usepackage{enumitem}
\usepackage{ifthen}
\usepackage{fullpage}

\usepackage[ruled]{algorithm2e} % For algorithms

\SetAlFnt{\small}
\SetAlCapFnt{\small}
\SetAlCapNameFnt{\small}
\SetAlCapHSkip{0pt}
\IncMargin{-\parindent}

\usepackage{booktabs} % For formal tables
\usepackage{xspace}
\usepackage[pagebackref=true,hidelinks]{hyperref}
\usepackage{accents} % for underbar
\usepackage{pgf} % for pgf figure; could be removed
\usepackage{xparse} % for DeclareDocumentCommand
\usepackage{bbm}
\usepackage{enumitem}
\usepackage[numbers]{natbib}
\usepackage{multirow}
\usepackage{graphicx}
\usepackage{subfigure}
\usepackage{wrapfig}
\usepackage{cleveref}

\graphicspath{ {figures/} }

\usepackage{fixme}
\fxsetup{multiuser,layout={inline,index}}
\fxusetheme{colorsig}
\FXRegisterAuthor{sc}{asc}{SC}
\FXRegisterAuthor{yt}{ayt}{YT}
\FXRegisterAuthor{ct}{act}{CT}

\newtheorem{theorem}{Theorem}[section]

\newtheorem{lemma}[theorem]{Lemma}
\newtheorem{corollary}[theorem]{Corollary}
\newtheorem{fact}[theorem]{Fact}

\theoremstyle{definition}
\newtheorem{definition}{Definition}[section]

\newenvironment{proofof}[1]{\vspace{0.1in}

\noindent{\em Proof of #1.}}{\hfill\qed}

\newenvironment{numberedtheorem}[1]{%
\renewcommand{\thetheorem}{#1}%
\begin{theorem}}{\end{theorem}\addtocounter{theorem}{-1}}

\newenvironment{numberedlemma}[1]{%
\renewcommand{\thetheorem}{#1}%
\begin{lemma}}{\end{lemma}\addtocounter{theorem}{-1}}

\newcommand{\argmax}{\operatorname{arg\,max}}
\newcommand{\argmin}{\operatorname{arg\,min}}

\newcommand{\prob}[2][]{\textrm{\bf Pr}\ifthenelse{\not\equal{}{#1}}{_{#1}}{}\!\left[#2\right]}
\newcommand{\expect}[2][]{\textrm{\bf E}\ifthenelse{\not\equal{}{#1}}{_{#1}}{}\!\left[#2\right]}

\newcommand{\dist}{\mathcal{D}}

\newcommand{\mech}{\mathcal{M}}
\newcommand{\alloc}[1][]{X\ifthenelse{\not\equal{}{#1}}{_{#1}}{}}
\newcommand{\paymt}[1][]{P\ifthenelse{\not\equal{}{#1}}{_{#1}}{}}
\newcommand{\price}[1][]{p\ifthenelse{\not\equal{}{#1}}{_{#1}}{}}

\newcommand{\swv}{{u_v}}
\newcommand{\rev}{{\normalfont\textsc{Rev}}}
\newcommand{\srev}{{\normalfont\textsc{SRev}}}
\newcommand{\brev}{{\normalfont\textsc{BRev}}}
\newcommand{\drev}{{\normalfont\textsc{DRev}}}
\newcommand{\sybilrev}{{\normalfont\textsc{SybilProofRev}}}
\newcommand{\adaptsybilrev}{{\normalfont\textsc{AdaptSybilProofRev}}}
\newcommand{\subaddrev}{{\normalfont\textsc{DetSybilProofRev}}}

\newcommand{\lotto}{\lambda}
\newcommand{\lottos}{\Lambda}
\newcommand{\gt}[1][T]{\tilde{g}_{#1,a}}

\newcommand{\core}{{\normalfont\textsc{CORE}}}
\newcommand{\tail}{{\normalfont\textsc{TAIL}}}

\newcommand{\bmax}{b_{\max}}
\newcommand{\bmin}{b_{\min}}

\newcommand{\setdist}{{\Pi}}
\newcommand{\fp}{{p_{\mathcal S, \beta}}}

\begin{document}
% Title portion. Note the short title for running heads 
\title{Buy-many mechanisms are not much better than item pricing\thanks{This work was supported
      in part by NSF grants CCF-1617505 and CCF-1704117. A preliminary
      announcement of this work is to appear at the ACM EC'19 conference.}}
\author{Shuchi Chawla
\thanks{
    {University of Wisconsin-Madison}. \tt{\{shuchi@cs.wisc.edu,
      yifengt@cs.wisc.edu, tzamos@wisc.edu\}.}}
\and Yifeng Teng$^\dagger$
\and Christos Tzamos$^\dagger$
}
\date{}

\maketitle
\thispagestyle{empty}

\begin{abstract}
%!TEX root =  ec-main.tex

Multi-item mechanisms can be very complex offering many different bundles to the buyer that could even be randomized. Such complexity is thought to be necessary as the revenue gaps between randomized and deterministic mechanisms, or deterministic and simple mechanisms are huge even for additive valuations.

We challenge this conventional belief by showing that these large gaps can only happen in restricted situations. These are situations where the mechanism overcharges a buyer for a bundle while selling individual items at much lower prices. Arguably this is impractical in many settings because the buyer can break his order into smaller pieces paying a much lower price overall. Our main result is that if the buyer is allowed to purchase as many (randomized) bundles as he pleases, the revenue of any multi-item mechanism is at most $O(\log n)$ times the revenue achievable by item pricing, where $n$ is the number of items. This holds in the most general setting possible, with an arbitrarily correlated distribution of buyer types and arbitrary valuations.

We also show that this result is tight in a very strong sense. Any family of mechanisms of subexponential description complexity cannot achieve better than logarithmic approximation even against the best deterministic mechanism and even for additive valuations. In contrast, item pricing that has linear description complexity matches this bound against randomized mechanisms.

\end{abstract}
%\newpage

\setcounter{page}{1}

%!TEX root =  main.tex
\section{Introduction}

It is well known that revenue optimal mechanisms can be complicated:
when a seller has more than one item to sell to a buyer, the seller
may price bundles of items rather than just the individual items;
better still the seller may offer random subsets of items, also called
lotteries; even more surprisingly, the seller may offer an infinitely
large menu of such options. A primary line of enquiry within
algorithmic mechanism design aims to establish simplicity versus
optimality tradeoffs: is it possible to obtain some fraction of the
optimal revenue via simple mechanisms, i.e. mechanisms that can be
described easily, understood easily, and are prevalent in practice?

To quantify this tradeoff let us introduce some standard notation. Let
\rev\ denote the optimal revenue obtained by using the most general
kind of mechanism---a menu of lottery pricings. Let \drev\ denote the
optimal revenue achieved by a deterministic mechanism---a menu of
bundle pricings. Two ``simple'' mechanisms have been studied
extensively in literature: item pricings, where each item is assigned
a price and the buyer can buy any subset at the sum of the constituent
prices; and bundle pricings, where every bundle is sold at a
single constant price. The optimal revenues achievable by item
pricings and bundle pricings are denoted \srev\ and \brev\
respectively. The goal then is to bound the ratios \rev/\drev, or
$\rev/\max(\srev,\brev)$, etc.

Briest et al.~\cite{BCKW-SODA10} were the first to show that revenue optimal
mechanism design exhibits a particularly diabolical curse of
dimensionality: whereas \rev/\srev\ is equal to 1 when the seller has
only one item to sell, the ratio becomes infinite with three items
even if the buyer is unit demand. Hart and Nisan~\cite{HN-EC13} improved this
to show that the ratio is infinite already with just two items. In
fact, if the buyer has additive values over two items, meaning that
the value of the bundle of two items is the sum over the individual
item values, the ratio of \rev/\drev\ can also be
infinite.\footnote{Note that for unit demand buyers \drev=\srev.} Even
focusing on just deterministic mechanisms, the situation is not much
better: the ratio of \drev\ to \srev\ can be as large as $\Omega(2^n/n)$
with $n$ items and additive values. Daskalakis in his 2014
survey~\cite{daskalakis2015multi} summarized the situation as such: 
``Multi-item auctions defy intuition.''

Further work along these lines suggested that the existence of a good
tradeoff depends on properties of the buyer's valuation -- both the
structure of the value function as well as the distribution from which
it is drawn. For example, two parallel lines of work investigating
unit demand buyers~\cite{CHK-07, CMS-10} and additive valuation
buyers~\cite{HN-EC12, LY-PNAS13, bilw-focs14} respectively established
that when the buyer's values for different items are drawn
independently, the larger of \srev\ and \brev\ is within a constant
factor of \rev. These results extend to more general settings with
subadditive valuations~\cite{rw-15, CM-16} as well as multiple
buyers~\cite{CHMS-10, yao-15, CM-16, CZ-17}, but continue to require
some degree of independence across individual item
values.\footnote{\cite{CMS-10}, \cite{rw-15}, and \cite{BDHS-15} allow
  item values to be correlated by being defined as linear functions
  over a common set of random variables, but the latter is required to
  be independent. There is one notable exception. 
  Psomas et al.~\cite{PSW-19} perform an investigation
  of smoothed complexity for revenue optimal mechanism design over
  correlated value distributions, where they studied the smoothed revenue 
  achievable after the value distribution is perturbed slightly in different ways.}

Our work presents an alternate view of the simplicity versus
optimality tradeoff. {\bf Our thesis is that the curse of dimensionality only
happens when buyers' actions are restricted.} 
Let us illustrate through an example adapted from Hart and
Nisan~\cite{HN-EC13}. Consider a seller with $n$ items and a single
buyer with additive values. Order all of the $2^n-1$ subsets of the
items in weakly increasing order of size, and let $S_i$ denote the
$i$th set in this ordering. The buyer's value function is picked
randomly from a set of $2^n-1$ different types. The $i$th type is
realized with probability proportional to $n^{-i}$ and values every
item in the set $S_i$ at $n^i/|S_i|$ and every item not in $S_i$ at
$0$. It is now straightforward to see that a mechanism $\mech$ that
offers the set $S_i$ at a price of $n^{i-1}$ extracts as revenue a
$1/n$ fraction of the buyer's total expected value, or
$\Omega(2^n/n)$---for every $i$, the type $i$ buyer extracts the most
utility by purchasing the set $S_i$. On the other hand, the buyer's
expected value is distributed according to the equal revenue
distribution, and so any bundle pricing extracts at most $O(1)$
revenue and any item pricing extracts at most $O(n)$. The implication
is that optimal deterministic mechanisms can obtain exponentially
larger revenue than simple ones.

We notice that $\mech$ disallows buyer from purchasing more than one
set of items offered by the seller. The single-buyer
mechanism design problem is a convenient abstraction for settings with
unlimited supply and multiple i.~i.~d. buyers. Upon finding a
(near-)optimal mechanism for the single-buyer setting, we can apply
the mechanism as-is in the latter setting, once for each
buyer. However, in that context, mechanisms such as $\mech$ offer
buyers opportunities for arbitrage. Consider, in particular, a set
$S_i$ of size $2$. $\mech$ sells both of the items in $S_i$
individually at a price no more than a $1/n$ fraction of the price of
$S_i$. A buyer of type $i$ can then participate in the mechanism
twice, purchasing the constituents of $S_i$ individually and paying
far less to the mechanism than before. In effect, $\mech$ is not
Sybil-proof.

Single-buyer mechanisms can without loss of generality be described as
menus of options where each option is a random allocation
(a.k.a. lottery) paired with a price. In this paper we study {\bf
  Sybil-proof} or {\bf buy-many} mechanisms where the buyer is allowed
to purchase any multi-set of menu options (of arbitrary
size).\footnote{When the menu contains lotteries, there is a slight
  distinction between whether the buyer can select a multi-set of
  options adaptively depending on outcomes of previous lotteries, or
  non-adaptively. Our results apply to both settings.} For
deterministic menus, Sybil-proofness simply means that the prices
assigned to different menu options are subadditive over subsets of
items. As a corollary, item pricing and bundle pricing are already
Sybil-proof. For general mechanisms, in fact, imposing the constraint
of Sybil-proofness greatly limits the extent to which mechanisms can
price discriminate between buyers of different types and, in
particular, disallows the gap examples of Briest et al. and Hart and
Nisan.\footnote{Babaioff et al.~\cite{babaioff2018optimal} previously
  observed that there can be a ``small positive constant'' revenue gap
  between Sybil-proof and optimal mechanisms, without bounding such a
  constant from below. Our results imply that this gap is unbounded.}

We can now again ask whether arbitrary, complicated Sybil-proof
mechanisms can obtain unboundedly larger revenue (or, even
exponentially or polynomially larger revenue, with respect to the
dimension $n$) relative to simple mechanisms. Our main result is that
they cannot:
\begin{theorem}\label{thm:ub}
  For any arbitrary distribution over arbitrary valuation functions, 
\[\frac{\sybilrev}{\srev} = O(\log n).\]
\end{theorem}

Here \sybilrev\ denotes the optimal revenue achievable through
(potentially randomized) Sybil-proof mechanisms. Briest et
al.~\cite{BCKW-SODA10} previously studied the gap between Sybil-proof
mechanisms\footnote{Briest et al. used the term ``buy-many setting''
  for what we call Sybil-proof mechanisms.} and item pricings for the
special case of unit-demand buyers and proved the same upper
bound. Our main contribution is to extend this result to arbitrary
valuation functions and distributions over valuations. Indeed we make
no assumptions whatsoever on the buyer's valuations other than that
they are monotone non-decreasing in the set of items
allocated---allocating extra items to the buyer never lowers his
value.

Can we do even better? We show that the above result is tight in a
very strong sense:
\begin{theorem}
  \label{thm:lb-intro}
  There exists a distribution over additive values for which no
  mechanism with description complexity at most $2^{o(n^{1/6})}$ can obtain a
  $o(\log n)$ fraction of the optimal deterministic Sybil-proof revenue.
\end{theorem}
  Observe that item pricings can be described using $O(n\log R)$ bits
  when values lie in the range $[1,R]$. We construct a distribution
  over additive values with $R<n$ and a deterministic subadditive
  pricing such that no mechanism that can be represented using
  $2^{o(n^{1/6})}$ bits can obtain a $o(\log n)$ fraction of the
  revenue of the subadditive pricing. We further show that for
  single-minded buyers, the gap cannot be improved in general even if
  the pricing we are comparing against is a submodular
  function. Briest et al. previously showed a similar result for
  unit-demand buyers, namely that the ratio $\frac{\sybilrev}{\srev}$
  can be $\Omega(\log n)$.

  \paragraph{Menu size.} Our results also have implications for the
  menu size complexity of optimal auctions. The menu size of an
  auction, defined as the number of different outcomes the seller
  offers to the buyer, has been studied extensively in literature as a
  measure of complexity for single-buyer mechanisms (see, e.g.,
  \cite{HN-EC13, DHN-14, BGN-17, Gonc-18}). One criticism of this
  notion of menu size is that some mechanisms can be described much
  more succinctly than indicated by their menu size, as is the case
  for item pricing. Hart and Nisan~\cite{HN-EC13} introduced the
  alternate concept of ``additive menu size'', namely the number of
  ``basic'' options a buy-many mechanism offers, and showed that even
  mechanisms with small additive menu size cannot capture a good
  fraction of the optimal revenue.  Our results show that allowing the
  buyer to purchase multiple options doesn't just allow a more
  succinct description of the mechanism, it also fundamentally changes
  the set of mechanisms available to the seller. In particular,
  buy-many mechanisms with even infinite additive menu size cannot
  capture any finite fraction of the overall optimal revenue \rev, as
  their revenue is bounded by \sybilrev.

  Our work calls for a new investigation of additive menu-size
  complexity of Sybil-proof mechanisms.  A natural question, for
  example, is whether one can always obtain a constant fraction of the
  optimal Sybil-proof revenue via mechanisms with finite additive menu
  size. If so, what menu size is necessary to obtain a $1-\epsilon$
  fraction of the optimal Sybil-proof revenue? Does it matter whether
  the buyer can select multiple options adaptively or non-adaptively?
  We leave these questions to future work.

  \paragraph{Our techniques.} Henceforth we will represent
  single-buyer mechanisms as pricing functions that assign a price to
  every possible (random) allocation. For Sybil-proof mechanisms, this
  price corresponds to the cheapest manner in which the buyer can
  acquire a (collection of) lottery(ies) that dominates the desired
  random allocation. For a pricing function $p$ and value function
  $v$, let $\rev_v(p)$ denote the 
  %price assigned to the lottery that a
  %buyer with value $v$ buys under $p$; this is the 
  revenue the mechanism $p$ obtains from a buyer with value $v$. When $v$ is drawn
  from a distribution $\dist$, the mechanism's expected revenue is
  $\expect[v\sim\dist] {\rev_v(p)}$. Our goal is to find for any given
  Sybil-proof pricing $p$ and distribution $\dist$, a ``simple''
  pricing $q$, such that $\expect[v\sim\dist] {\rev_v(q)}$ is comparable to
  $\expect[v\sim\dist] {\rev_v(p)}$. 

  As a first attempt towards this goal, we ask whether $p$ can be
  ``point-wise'' approximated by $q$. That is, does there exist a
  small $c>1$ such that for all random allocations $\lambda$,
  $q(\lambda)/c \le p(\lambda)\le q(\lambda)$? It turns out that for
  adaptively Sybil-proof pricings $p$,\footnote{Non-adaptive
    Sybil-proofness requires some extra work in dealing with lotteries
    that have valuable items at very very low probabilities of
    allocation.} with $q$ being an additive function, taking $c=n$
  suffices. Since our eventual goal is to obtain an approximation to
  revenue in expectation over a given value distribution, rather than
  point-wise over each possible lottery purchased, it is reasonable to
  expect that a scaling-type argument would provide an $O(\log c)$
  approximation. In particular, let $\alpha$ be a random power of $2$
  between $1/c$ and $1$, and let $\alpha q$ denote the pricing where
  every price in $q$ is scaled by the factor $\alpha$. Then it holds
  that for any $\lambda$, with probability $1/\log(c)$,
  $\alpha q(\lambda)$ is within $[1/2,1]$ times $p(\lambda)$. The
  implication is that a buyer purchasing $\lambda$ under $p$ would
  still be interested in purchasing $\lambda$ at the cheaper price
  offered by $\alpha q$, while paying at least half of what he was
  paying under $p$. However, there is a fallacy in this argument: a
  buyer that purchases $\lambda$ under $p$ may switch to purchasing a
  much cheaper allocation when offered the pricing $\alpha q$. Our
  main technical contribution is in dealing with the buyer's incentives
  to argue that even if the buyer switches allocations, the resulting
  revenue is still significant large. 

  Formally we prove the following theorem, that may be of independent
  interest. The theorem states that if a pricing function $q$ {\em
    point-wise $c$-approximates} a pricing function $p$, then a
  particular (random) scaling of $q$ obtains an {\em
    $O(\log c)$-approximation in revenue} to $p$ with respect to {\em
    any} arbitrary distribution over valuation functions.

  \begin{theorem}
  \label{thm:general}
    For any $c>1$, let $p$ and $q$ be any two pricing functions
    satisfying $q(\lambda)/c \le p(\lambda) \le q(\lambda)$ for all
    random allocations $\lambda\in \Delta(2^{[n]})$. Then there exists a
    distribution over ``scaling factors'' $\alpha\in [1/2c,1]$ such that
    for any valuation function $v$, $\operatorname{E}_\alpha[\rev_v(\alpha
    q)]\ge \frac{1}{2\log(2c)} \rev_v(p)$.
  \end{theorem}

  Theorem~\ref{thm:general} shows that approximating a pricing function pointwise
  suffices to obtain a good approximation to the revenue.
  
  Our lower bounds are obtained by considering a weaker notion of approximation
  of a pricing function $p$ using another function $q$ that exactly captures the revenue
  tradeoff over many natural valuation classes. It only requires 
  the approximation be accurate in expectation over the ``demand distribution'',
  i.e. the distribution over sets that are
  bought when the function $p$ is offered as a price menu to the
  buyer. We say that a pricing function $q$, \emph{$c$-approximates from below} the pricing function $p$  over the demand distribution $\Pi$ if, 
  \begin{align}
    \label{eq:approximation-in-expectation}
    \expect[S\sim\Pi]{q(S) \mathbbm{1}_{q(S)\le p(S)}} \ge \frac 1 c \expect[S\sim\Pi]{p(S)} 
  \end{align}

%  Our lower
%  bounds are constructed essentially by viewing the problem of
%  approximating the revenue of one pricing function using another in
%  the following manner. Let $p:2^{[n]}\rightarrow \R$ be the
%  subadditive function we are trying to approximate. Let $\Pi$ denote
%  the ``demand distribution'', or the distribution over sets that are
%  bought when the function $p$ is offered as a price menu to the
%  buyer. Then, the expected revenue of $p$ can be written as
%\begin{align}
%  \label{eq:simplerev-f}
%  \expect[S\sim\Pi]{p(S)}
%\end{align}
%Let $q:2^{[n]}\rightarrow \R$ be another pricing function that we are
%trying to construct. Then, without any further information about the
%buyer's value distribution (and ignoring incentives for a moment), 
%it turns out that for any set $S$ with $q(S)\le p(S)$, $q$ obtains a
%revenue of $q(S)$ from buyers that purchase $S$ under $p$, and for any
%set $S$ with $q(S)> p(S)$, $q$ obtains a revenue of $0$. Therefore, a
%conservative estimate for the revenue obtained by $q$ is:
%\begin{align}
%  \label{eq:simplerev-g}
%\expect[S\sim\Pi]{q(S) \mathbbm{1}_{q(S)\le p(S)}}
%\end{align}

Note that the expectation in the right hand side of \eqref{eq:approximation-in-expectation} is
exactly the revenue of $p$, while the expectation in the left hand side is a proxy
for the revenue of $q$. It assumes that the demand distribution over sets remains the same
but a set is bought only if its price under $q$ is not higher than its price under $p$.

In fact, there is a distribution over single-minded
buyers\footnote{Single-minded buyers are interested in buying a specific
  subset $S$ of items at some value.} for
which the expected revenues of $p$ and $q$ are exactly equal to the
expressions above: we pick a set $S$ from the distribution $\Pi$ and
assign a value of $p(S)$ to every superset of $S$, including itself,
and a value of $0$ everywhere else.

Our lower bounds show that under this weaker notion of approximation,
functions $q$ with subexponential description complexity cannot
obtain better than $O(\log n)$ approximation, even for submodular pricing functions
$p$. This directly implies that no simple
mechanism can get better than logarithmic approximation to the optimal revenue 
for single minded-buyers (Theorem~\ref{thm:lb-singleminded}) and also extends
to the case of additive buyers (Theorem~\ref{thm:lb-additive}).

%Now we may ask, for a given function $p$, does there exist a simple
%$q$ for which the expression in \eqref{eq:simplerev-g} approximates
%the expression in \eqref{eq:simplerev-f}? As we discussed above, the
%answer is yes with a gap of $O(\log n)$, but we cannot hope for a
%smaller gap in general. 

%In Appendix~\ref{sec:better} we examine whether better upper bounds
%can be obtained in special cases.

While in general better-than-logarithmic approximation with simple mechanisms
is not possible even for simple valuations, in Appendix~\ref{sec:better} 
we offer improved upper bounds
in special cases where the demand distribution has additional structure.

% Can we obtain a better bound in special cases? We show for
% single-minded buyers that if the demand distribution $\Pi$ is a
% product distribution then the better of item pricing and bundle
% pricing obtains a constant factor approximation to the revenue of
% $p$. Our approach for proving this result is very similar to that of
% Rubinstein and Weinberg~\cite{rw-15} who show a constant upper bound
% on the ratio of \rev\ to $\max(\srev,\brev)$ when the buyer's value
% function is subadditive over independent item values. Despite the
% similarity in techniques, our setting is quite different from that of
% Rubinstein and Weinberg. In our setting, the buyer's value function is
% single-minded (and in particular, not subadditive), the optimal
% pricing we compare against is subadditive (not arbitrary), and the
% independence is across the items demanded by the buyer (and not the
% item values). Further exploring the simplicity versus optimality tradeoff in
% terms of demand distributions is another fruitful direction for
% future work.

\section{Notation and definitions}
\label{sec:prelim}

We study the following single-buyer mechanism design problem. A seller has $n$ heterogeneous items to sell to a single buyer. The buyer's type is given by a valuation function $v$ that assigns non-negative values to every set of items: $v:2^{[n]}\to \mathcal{R}_0^+$. Values are monotone, meaning that for any $S$ and $T$ with $S\subset T\subseteq [n]$, $v(S)\le v(T)$. The buyer's type is drawn from an arbitrary known distribution $\dist$ over the set of all possible valuation functions.

Any selling mechanism can be described as a menu of options, each of which assigns a price to a random allocation or lottery. Let $\Delta=\Delta(2^{[n]})$ denote the set of all probability distributions over sets of items and $\lotto\in\Delta$ denote a ``lottery'' or random allocation. We describe a mechanism using a pricing function $p:\Delta \to \mathcal{R}_0^+$; The price assigned to a lottery $\lotto\in\Delta$ is then given by $p(\lotto)$.\footnote{\label{ftnote}Since we are not investigating menu size, we will assume that the pricing assigns a price to every lottery. It is easy to extend a partial pricing to a complete one: we assign to every lottery the price of the cheapest option that dominates it, or a price of infinity if no such option exists. For Sybil-proof pricings, we define the price for some lottery $\lambda$ as being the cheapest way to assemble a random allocation $\lambda'$ that first-order stochastically dominates $\lambda$. See below for a formal definition of dominance among lotteries.}

If a buyer with valuation $v$ buys a lottery $\lotto\in\Delta$ at price $p(\lotto)$, her utility from the purchase is given by
\[u(v,p,\lotto):=\expect[S\sim\lotto]{v(S)}-p(\lotto).\]
If a buyer with valuation $v$ purchases a multiset $\lottos=\{\lotto_1, \lotto_2, \cdots, \lotto_k\}$ of lotteries with price function $p$, her utility from the purchase is given by
\[u(v,p,\lottos):=\expect[S_i\sim\lotto_i \forall i\in {[k]}]{v\left(\bigcup_i S_i\right)}-\sum_{i\in [k]}p(\lotto_i).\]
We sometimes use $v(\lotto)$ as shorthand for $\expect[S\sim\lotto]{v(S)}$ and $v(\lottos)$ likewise for the value assigned to the union of sets drawn from a multiset $\lottos$ of lotteries. 

\paragraph{Sybil-proofness and optimal revenue.} We say that a lottery $\lotto$ dominates another lottery (or a multiset of lotteries) $\lotto'$ if there exists a coupling between a random draw $S$ from $\lotto$ and a random draw (or union of draws) $S'$ from $\lotto'$ such that $S$ is a superset of $S'$. We are now ready to define Sybil-proofness.

\begin{definition}
A mechanism or pricing $p:\Delta \to \mathcal{R}_0^+$ is {\bf Sybil-proof} if for {\em every} multiset $\lottos$ of lotteries there exists a single lottery $\lotto$ dominating it that is no more expensive: $p(\lotto)\le \sum_{\lotto'\in\lottos} p(\lotto')$.  
\end{definition}

Observe that if a lottery $\lotto$ dominates another lottery (or a multiset of lotteries) $\lotto'$, then any buyer with a monotone valuation function obtains higher expected value from $\lotto$ than from $\lotto'$. We therefore get the following observation:

\begin{fact}
  Given any Sybil-proof pricing $p$, for any buyer type $v$ it is optimal for the buyer to purchase a single lottery $\lotto = \argmax_{\lotto\in\Delta} (v(\lotto)-p(\lotto))$.
\end{fact}

Given a Sybil-proof pricing $p$, we use $\lotto_p(v) := \argmax_{\lotto\in\Delta} (v(\lotto)-p(\lotto))$ to denote the lottery purchased by a buyer with value function $v$, and $u(v,p)$ as the corresponding utility achieved. For convenience, we overload notation and use $p(v) := p(\lotto_p(v))$ to denote the price paid by the buyer. Given a distribution $\dist$ over buyer types, we write $\rev_\dist(p) := \expect[v\sim\dist] {p(v)}$ as the revenue of the pricing $p$. The optimal Sybil-proof revenue for distribution $\dist$ is given as follows; we drop the subscript $\dist$ when it is clear from the context.
\[\sybilrev_\dist := \max_{\text{Sybil-proof pricings } p} \rev_\dist(p). \]

\paragraph{Adaptive Sybil-proofness.} Our definition of Sybil-proofness guards against buyers that purchase multiple options from the given menu and receive a random allocation for each. The buyer can perform even better if these menu options are selected sequentially and adaptively---that is, if the buyer observes the instantiation of each random allocation before deciding whether and what to purchase next. Guarding against such an adaptive buyer places a further restriction on the prices the mechanism can charge. Let $\mathcal A$ denote a buying strategy, that is, an adaptive sequence of lotteries. Let $\lottos_{\mathcal A}$ denote the (random) sequence of lotteries bought in $\mathcal A$. As before, we say that $\mathcal A$ is dominated by a lottery $\lotto$ if there exists a coupling between a random draw $S$ from $\lotto$ and a random union of draws $S'$ from $\lottos_{\mathcal A}$ such that $S$ is a superset of $S'$. We can now define a more restrictive definition of Sybil-proofness:

\begin{definition}
A mechanism or pricing $p:\Delta \to \mathcal{R}_0^+$ is {\bf Adaptively Sybil-proof} if for {\em every} adaptive buying strategy $\mathcal A$ there exists a single lottery $\lotto$ dominating it that is cheaper: $p(\lotto)\le \expect[\lottos_{\mathcal A}]{\sum_{\lotto'\in\lottos_{\mathcal A}} p(\lotto')}$.  
\end{definition}

Of course there is no way for the seller to enforce whether buyers can make purchasing decisions adaptively or non-adaptively, so it is natural to study mechanisms satisfying adaptive Sybil-proofness. Competing against optimal adaptively Sybil-proof mechanisms makes some of our arguments easier. However, all of our positive results also apply to the less restrictive notion of (non-adaptive) Sybil-proofness.

%  Nevertheless, all of our positive results apply for the more restrictive notion of (non-adaptive) Sybil-proofness. We define the optimal Adaptively Sybil-proof revenue for distribution $\dist$ as follows.
% \[\adaptsybilrev_\dist := \max_{\text{Sybil-proof pricings } p} \rev_\dist(p). \]

\paragraph{Deterministic mechanisms.} Deterministic mechanisms price only deterministic sets (a.k.a. bundles) of items---$p:2^{[n]}\to \mathcal{R}_0^+$. A deterministic pricing $p$ is Sybil-proof if and only if it is monotone and subadditive, that is, for any $S_1\subset S_2\subseteq [n]$, $p(S_1)\le p(S_2)$, and for any set of bundles, $\mathcal{T}\subseteq 2^{[n]}$, the prices of the union of bundles is no more than the sum of individual bundle prices: $\sum_{S\in \mathcal{T}} p(S) \ge p(\cup_{S\in \mathcal{T}}\, S)$. The optimal deterministic Sybil-proof revenue for distribution $\dist$ is given as follows.
\[\subaddrev_\dist := \max_{\substack{\text{Deterministic monotone}\\ \text{subadditive pricings } p}} \rev_\dist(p).\]

\paragraph{Simple pricings.} An item pricing is a deterministic additive pricing: $p(S) = \sum_{i\in S} p(\{i\})$ for all $S\subseteq [n]$. A bundle pricing is a constant pricing that assigns the same price to every set: $p(S)=p([n])$ for all $S\subseteq [n]$. Observe that item pricings and bundle pricings are always Sybil-proof. We use \srev\ and \brev\ to denote the optimal revenue achievable using item pricings and bundle pricings respectively (over an implicit distribution $\dist$).

\section{Approximation via item pricing}
\label{sec:upperbound}

In this section we present our main upper bound, namely that the ratio between $\sybilrev_\dist$ and $\srev_\dist$ is bounded by $O(\log n)$ for any value distribution $\dist$. This result is based on ideas that are present implicitly in the work of Briest et al.~\cite{BCKW-SODA10} for unit-demand buyers. We formalize and extend these ideas to arbitrary valuations. 

Our argument proceeds in two parts. First, we show that in order to approximate the expected revenue of a pricing $p$ with a pricing $q$ within some factor $\beta$, it suffices to obtain a point-wise $2^{O(\beta)}$-approximation of $p$ via $q$. Then we show that adaptively Sybil-proof pricings $p$ can be point-wise $n$-approximated by additive pricings. This provides a bound on the gap between adaptive $\sybilrev_\dist$ and $\srev_\dist$. In the last part of this section, we extend our result to non-adaptively Sybil-proof pricings. 

\subsection{Point-wise approximation implies revenue approximation}

Suppose that we have two pricing functions $p$ and $q$ that are close on every possible (random) allocation. In particular, for some $c>1$, we have $q(\lambda)/c\leq p(\lambda)\leq q(\lambda)$ for every random allocation $\lambda\in\Delta$. We say that $q$ point-wise $c$-approximates $p$. We will now prove Theorem~\ref{thm:general}, namely that a point-wise $c$-approximation implies an $O(\log c)$-approximation in revenue. 

 \begin{numberedtheorem}{\ref{thm:general}}
    For any $c>1$, let $p$ and $q$ be any two pricing functions
    satisfying $q(\lambda)/c \le p(\lambda) \le q(\lambda)$ for all
    random allocations $\lambda\in \Delta(2^{[n]})$. Then there exists a
    distribution over ``scaling factors'' $\alpha\in [1/2c,1]$ such that
    for any valuation function $v$, $\operatorname{E}_\alpha[\rev_v(\alpha
    q)]\ge \frac{1}{2\log(2c)} \rev_v(p)$.
  \end{numberedtheorem}

Our argument considers a suite of pricings $\{\alpha q\}$ defined over a range of scale factors $\alpha\in [1/2c,1]$. We want to argue that this suite of pricings collectively obtains good revenue from {\em every} buyer type. At the heart of our argument is the following observation: for any buyer type, if the buyer obtains much larger utility at low prices ($\frac 1{2c} q$) than at high prices ($q$), we can capture a good fraction of this difference in utilities as revenue by picking a scaling factor $\alpha$ from an appropriate distribution independent of the buyer's type. We formalize this observation as Lemma~\ref{lem:util-as-rev} below. We then argue that if $q$ point-wise approximates $p$, then this difference of utilities is proportional to the revenue obtained by $p$ from the buyer.

\begin{figure}[htbp]
        \centering
        \includegraphics[width=0.95\textwidth]{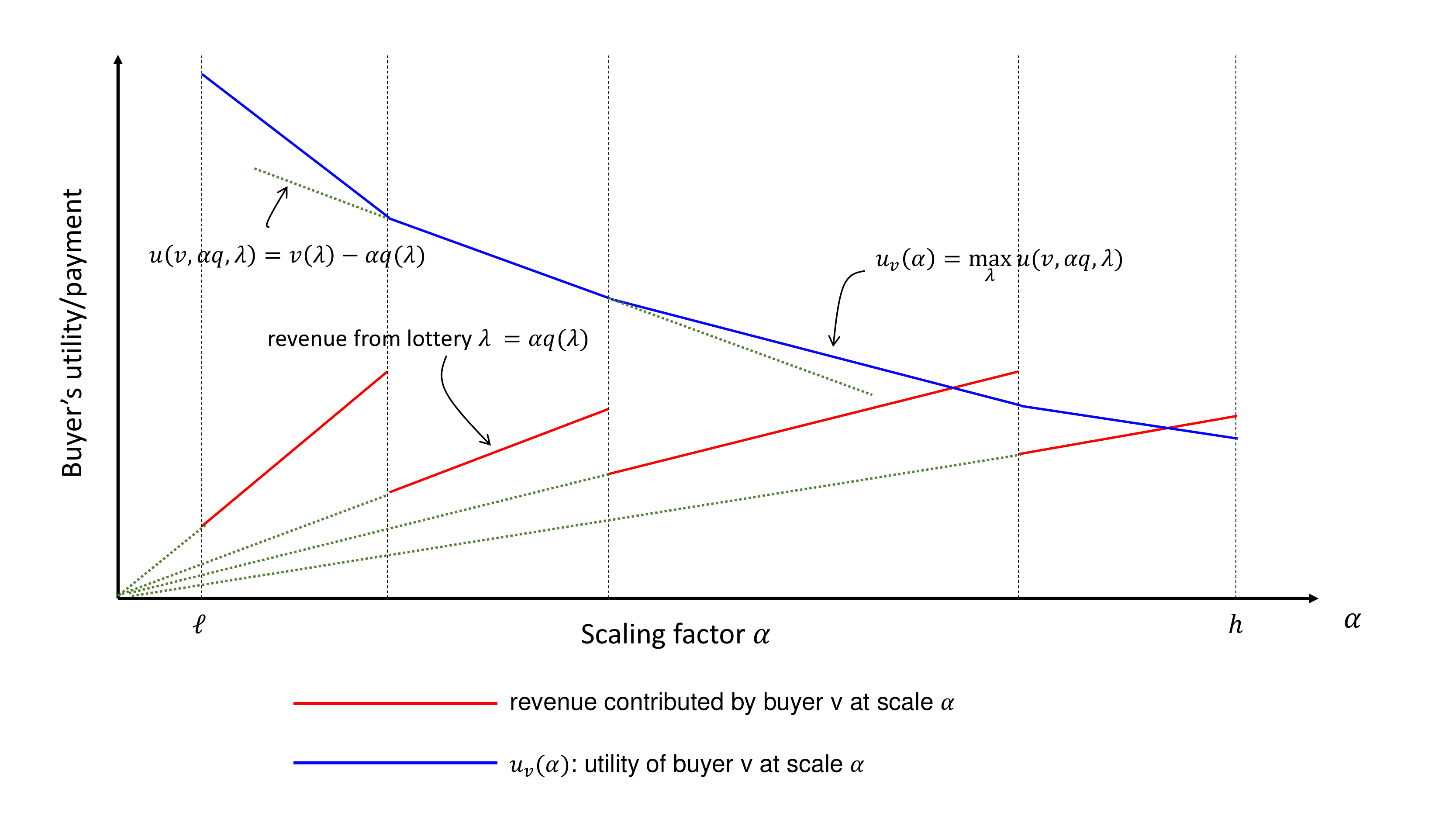}
        \caption{\label{fig:example} \small Utility and revenue as function of the scaling factor $\alpha$. Our goal is to find a distribution on $\alpha$ such that the expected revenue (red curve) is proportional to the change in utility (blue curve).}
    \end{figure}

\begin{lemma}
  \label{lem:util-as-rev}
  For any pricing $q$ and any $0<\ell\le h$, let $\alpha$ be drawn from $[\ell,h]$ with density function $\frac 1{\alpha\log(h/\ell)}$. Then, for any valuation function $v$,
\[\operatorname{E}_\alpha[\rev_v(\alpha q)]\ge \frac{u(v,\ell q)-u(v,h q)}{\log(h/\ell)}.\]
\end{lemma}
To understand the lemma, consider starting at $\alpha=\ell$ and gradually increasing the scaling factor and, correspondingly, prices. As the prices increases, the buyer's utility from purchasing his favorite bundle weakly decreases. As long as the buyer continues to buy the same lottery, this decrease in utility is captured as revenue. At certain price points the buyer switches to buying a different lottery, causing the revenue to drop discontinuously. Other than these discontinuities, however, the buyer's loss in utility is captured exactly as increase in revenue (see Figure~\ref{fig:example}). This observation allows us to relate the gradient of the revenue as a function of the scaling factor $\alpha$ to the gradient of the utility. Our next observation is that because revenue depends linearly on the scaling factor $\alpha$ (except at the break points), the revenue is proportional to its gradient. Then by picking an appropriate distribution on $\alpha$, we can capture exactly this gradient as expected revenue. We now present this argument formally.

\begin{proof}
Consider a buyer with value function $v$. Let $\swv(\alpha)$ denote the utility the buyer derives when offered the pricing $\alpha q$, and $\lambda_\alpha$ denote the corresponding lottery purchased:
\begin{align*}
\swv(\alpha) & = \max_{\lambda\in \Delta} \, (v(\lambda)-\alpha q(\lambda))\\
\lambda_\alpha & = \argmax_{\lambda\in \Delta} \, (v(\lambda)-\alpha q(\lambda)).
\end{align*}
By the envelope theorem it holds that $\swv'(\alpha) = \frac{\textrm{d}}{\textrm{d}\alpha}(v(\lambda_\alpha)-\alpha q(\lambda_\alpha))=-q(\lambda_\alpha)$. Therefore, the seller's revenue from this buyer is given by 
\[\alpha q(\lambda_\alpha) = -\alpha\swv'(\alpha). \]

Now, consider picking $\alpha$ from the equal revenue distribution over $[\ell,h]$ with density function $1/(\alpha\log (h/\ell))$, and offering the buyer the pricing $\alpha q$. Then, the expected revenue from this buyer is:
\begin{align*} 
\int_{\ell}^h  \alpha q(\lambda_\alpha) \cdot \frac{1}{\alpha\log (h/\ell)} d\alpha & =  \frac{-1}{\log (h/\ell)}\int_{\ell}^h \swv'(\alpha) d\alpha 
 = \frac{\swv(\ell)-\swv(h)}{\log (h/\ell)}.
\end{align*}  
\end{proof}

\noindent
We are now ready to complete the proof of Theorem~\ref{thm:general}.

\begin{proofof}{Theorem~\ref{thm:general}}
  Fix a buyer valuation $v$, and let $h=1$ and $\ell=1/2c$. We will bound the utility the buyer derives under pricings $\ell q$ and $hq$ in terms of his utility under $p$.
 \begin{equation*}
u(v,\ell q) = \max_{\lotto\in\Delta} \left(v(\lotto)-\frac{1}{2c} q(\lotto)\right) \ge \max_{\lotto\in\Delta} \left(v(\lotto)-\frac 12 p(\lotto)\right) \ge u(v,p) + \frac 12 p(v).
\end{equation*}
On the other hand, 
\begin{align*}
u(v,hq) = \max_{\lotto\in\Delta} \left(v(\lotto)-q(\lotto)\right)\le \max_{\lotto\in\Delta} \left(v(\lotto)-p(\lotto)\right) = u(v,p).
\end{align*}
We can now apply Lemma~\ref{lem:util-as-rev} to obtain a revenue of at least $\frac 1{2\log (2c)} p(v)$ from the buyer.
\end{proofof}

\subsection{Upper bound for adaptively Sybil-proof pricings}
\label{sec:ub-det}
As a warm-up to our main theorem, we show how Theorem \ref{thm:general} can be directly applied to get an upper bound on $\sybilrev_\dist/\srev_\dist$ for adaptively Sybil-proof pricings. We first briefly sketch the proof for deterministic Sybil-proof pricings. Recall that deterministic Sybil-proof pricings are monotone subadditive functions. Let $p$ be any such pricing. Then, by defining $q$ to be identical to $p$ over singletons ($q(\{i\})=p(\{i\})$ for all $i\in [n]$) and extending it additively to bundles ($q(S)=\sum_{i\in S} q(\{i\})$), we observe on the one hand that $q$ is no smaller than $p$ on any set $S$. On the other hand, $p(S)$ is at least as large as the price of the most expensive singleton item in $S$, which is at least as large as the average price of an item in $S$, or $q(S)/n$. Therefore, $q$ point-wise $n$-approximates $p$. We can therefore apply Theorem \ref{thm:general} to obtain an $O(\log n)$ approximation to the expected revenue of $p$ under any value distribution. We now formalize this argument for randomized adaptively Sybil-proof pricings.

% \begin{theorem}
% \label{thm:ub-det}
% For any distribution $\dist$, we have $\subaddrev_\dist\le O(\log n)\srev_\dist$.
% \end{theorem}

% \begin{proof}
%   Let $p$ be the subadditive pricing that achieves revenue $\subaddrev_\dist$. Let $q$ be an item pricing defined by $q(\{i\})=p(\{i\})$ for all $i\in [n]$. For any lottery $\lotto\in\Delta$, observe that $p(\lotto)=\sum_{S\subseteq[n]}\lambda(S)p(S)$, $q(\lotto)=\sum_{S\subseteq[n]}\lambda(S)\sum_{i\in S}q(\{i\})$. For all $S\subseteq [n]$, the subadditivity of $p$ implies
%   \[ \frac 1n q(S) \le \min_{i\in S}p(\{i\})\le p(S) \le q(S). \]
%   Thus for every $\lambda\in\Delta$, $\frac{1}{n}q(\lambda)\leq p(\lambda)\leq q(\lambda)$. The theorem immediately follows by taking $c=n$ in Theorem \ref{thm:general}.
% \end{proof}

% Similarly we prove the upper bound for adaptively Sybil-proof pricings by applying Theorem \ref{thm:general}.

\begin{theorem}
\label{thm:ub-adapt}
For any distribution $\dist$, we have $\adaptsybilrev_\dist\le O(\log n)\srev_\dist$.
\end{theorem}
\begin{proof}
  Let $p$ be the adaptive Sybil-proof pricing that achieves revenue $\adaptsybilrev_\dist$. We again begin by defining base prices for each of the items. Informally, these are the minimum prices the buyer needs to pay in expectation to obtain item $i$ with certainty:
\[ q(\{i\}) = \min_{\lotto\in\Delta}\, \frac{p(\lotto)}{\Pr[i\in \lotto]} \]
where we write $\Pr[i\in\lotto]$ as shorthand for $\Pr_{S\sim\lotto}[i\in S]$. Let $\lotto_i$ denote the lottery that defines the price for item $i$: $\lotto_i = \argmin_{\lotto\in\Delta}\, \frac{p(\lotto)}{\Pr[i\in \lotto]}$. For every $\lotto\in\Delta$, $q(\lotto)=\sum_{S\subseteq[n]}\lambda(S)\sum_{i\in S}q(\{i\})$.

Observe that for any lottery $\lotto\in\Delta$, the adaptive buyer can draw a set $T$ from $\lotto$ and purchase this set by purchasing for every $i\in T$ the lottery $\lotto_i$ repeatedly until he obtains the item. The expected price paid by the buyer in this strategy is precisely $\expect[T\sim\lotto]{q(T)} = \sum_{i\in [n]} (q(\{i\})\Pr[i\in\lotto])=q(\lotto)$. Since $p$ is adaptively Sybil-proof, we have $p(\lotto)\le q(\lotto)$.

On the other hand, by the definition of $q$, $p(\lotto)\ge q(\{i\}) \Pr[i\in\lotto]$ and therefore, summing over all $i\in [n]$ and dividing by $n$, we have $p(\lotto)\ge \frac 1n q(\lotto)$. Taking $c=n$ in Theorem \ref{thm:general} finishes the proof.
\end{proof}

\subsection{Upper bound for general randomized Sybil-proof pricings}
\label{sec:ub-rand}

We will now prove our main theorem:
\begin{numberedtheorem}{\ref{thm:ub}}
  For any arbitrary distribution over arbitrary valuation functions, 
\[\frac{\sybilrev}{\srev} = O(\log n).\]
\end{numberedtheorem}

% \begin{theorem}
% For any distribution $\dist$, we have $\sybilrev_\dist\le O(\log n)\srev_\dist$.
% \end{theorem}

As in our argument for the adaptive setting, consider any Sybil-proof pricing $p$ and let us define the pricing $q$ as: 
\[ q(\{i\}) = \min_{\lotto\in\Delta}\, \frac{p(\lotto)}{\Pr[i\in \lotto]} \,\,\forall i\in [n] \qquad \text{ and } \qquad q(S) = \sum_{i\in S} q(\{i\})\,\,\forall S\subseteq [n].\]
Our goal, as before, is to argue that $q$ point-wise approximates $p$. It is straightforward to argue that $p(\lotto) \ge \frac 1n q(\lotto)$ for all $\lotto\in \Delta$. However, it is no longer necessarily true that $p(S)\le q(S)$ for all sets $S$ of items. This is because acquiring the set $S$ non-adaptively with certainty under $p$ may cost much more than $q(S)$. This in turn implies that the buyer's utility under $q$ may be larger than his utility under $p$. Indeed no matter how large of a scaling factor $h$ we pick, it is tricky to directly argue that $u(v,hq)\le u(v,p)$. Instead, we will argue that if the set $S$ the buyer wants to purchase under $hq$ is too expensive to purchase under $p$, this happens because some high value item in $S$ can only be bought with very low probability in $p$. We will use such items as the basis to construct a different additive pricing that obtains a good fraction of the revenue of $p$. We now formalize this argument.

\begin{proof}
  Let $p$ be the Sybil-proof pricing that achieves revenue $\sybilrev_\dist$. Let $\ell=1/2n$, $h=4n$, and define the pricing $q$ as above. Let $\alpha$ be picked from the range $[\ell,h]$ with density function $1/(\alpha \log (h/\ell))$. Then for any valuation function $v$, it is straightforward to observe that $u(v,\ell q)\ge u(v,p)+p(v)/2$. We will now focus on bounding $u(v,hq)$.

% acquiring the lottery $\lotto$ non-adaptively may cost 

% Directly apply Theorem \ref{thm:general} to non-adaptive buyers requires some modification because it is no longer necessarily true that $p(\lotto)\le q(\lotto)$. Let us first reiterate the parts of the argument that work as before. We define the pricing $q$ as: 
% \[ q(\{i\}) = \min_{\lotto\in\Delta}\, \frac{p(\lotto)}{\Pr[i\in \lotto]} \,\,\forall i\in [n] \qquad \text{ and } \qquad q(S) = \sum_{i\in S} q(\{i\})\,\,\forall S\subseteq [n].\]
% Let $\alpha$ be picked from the range $[1/2n,4n]$ with density function $1/(\alpha \log (8n^2))$. Define $g_\alpha$, $S_\alpha$, $\swv(\alpha)$, and $\lotto_i$ as before. We then have:
% \begin{align}
% \notag  p(\lotto)& \ge \frac 1n q(\lotto) \\
% \label{eq:integral}  \expect[\alpha]{g_\alpha(v)} & = \frac{\swv(1/2n)-\swv(4n)}{\log (8n^2)} \\
% \intertext{and,}
% \notag  \swv(1/2n) & = \max_{S\subseteq [n]} \left(v(S)-\frac 1{2n} q(S)\right) \ge v(\lotto_p(v))-\frac 1{2n} q(\lotto_p(v)) \\
% \label{eq:rl} & \ge v(\lotto_p(v))-\frac 1{2} p(\lotto_p(v)) = u(v,p) + \frac 12 p(v). 
% \end{align}
% where $\lotto_p(v)$ is the lottery bought by the buyer in pricing $p$.

% Define $h=4n$. It remains to bound the buyer's utility in $g_h$, namely $\swv(h)$. 

Let $T_h$ be the set purchased by the buyer when offered the pricing $hq$. Then $u(v,hq)=v(T_h)-hq(T_h)$. We want to bound this utility in terms of the utility the buyer gets under the pricing $p$, so let us consider how much it would cost the buyer to acquire $T_h$ under $p$. In particular, fix some number $m$ and suppose that the buyer purchases a multiset $\lottos_m$ that contains $m_i=\lceil\frac{m}{\Pr[i\in \lotto_i]}\rceil$ copies of $\lotto_i$ for all $i\in T_h$. Then, the probability that some $i\in T_h$ does not belong to the random allocation drawn from this multiset is at most $(1-\Pr[i\in \lotto_i])^{m_i}< 2^{-m}$. Accordingly, the probability that $T_h$ is not a subset of the random allocation drawn from $\lottos_m$ is at most $n 2^{-m}$. The total price of $\lottos_m$ is $\sum_{i\in T_h} \lceil\frac{m}{\Pr[i\in \lotto_i]}\rceil \, p(\lotto_i) \le (m+1)\sum_{i\in T_h} q(\{i\}) = (m+1)q(T_h)$. We therefore have:
\[ 
u(v,p) \ge u(v,p,\lottos_m) \ge (1-n 2^{-m}) v(T_h) - (m+1)q(T_h) \quad \text{for all } m.
\]

Now, let $k$ be defined such that $n 2^{-k}v(T_h) = q(T_h)$, that is, $k = \log(nv(T_h)/q(T_h))$. Then we get:
\begin{align}
\notag u(v,p) & \ge v(T_h) - (k+2)q(T_h) & \text{and,} \\
\label{eq:rh} u(v,hq) & \le u(v,p) + (k+2-h) q(T_h). & 
\end{align}
Now, if $k+2\le h$, then we already have the bound we desire. So, for the remainder of the proof, assume that $k>h-2\ge 3n$. We will construct a different item pricing to recover the quantity $(k+2-h)q(T_h)<kq(T_h)$ as revenue from the buyer with value $v$.

For any $T\subseteq [n]$ and positive integer $a$, the (uniform) item pricing $\gt$ is defined as follows: $\gt(S) = 2^{a+n} q(T) |S|$ for all $S\subseteq [n]$. Now consider a buyer with value function $v$ and with $k = \log(nv(T_h)/q(T_h))$. When offered the pricing $\gt[T_h]$ with $a< k-n-2\log n$, this buyer obtains a utility of at least $v(T_h)-\gt[T_h](T_h) \ge v(T_h) - 2^{a+n}nq(T_h)> v(T_h)-\frac 1n2^kq(T_h)= 0$. Therefore, the buyer buys at least one item under this pricing and pays at least $2^{a+n}q(T_h)$. In other words,
\[
\gt[T_h](v) \ge 2^{a+n}q(T_h) \quad \text{for } a< k-n-2\log n.  
\] 
Let $T$ be drawn from the uniform distribution over $2^{[n]}$ and $a$ be drawn from the geometric distribution with mean 2, that is, $\Pr[a=x]=2^{-x-1}$ for every $x\in\mathbb{N}$. Then, we get:
\begin{align}
\notag \expect[T,a]{\rev_v(\gt)} & \ge \frac 1{2^n} \sum_{a=0}^{k-n-2\log n-1} \frac 1{2^{a+1}}2^{a+n}q(T_h) \\
\notag & = \frac 12 (k-n-2\log n-1) q(T_h)\\
\label{eq:alt-pricing} & \ge \frac 14 kq(T_h) \quad \text{for } k\ge h-2.
\end{align}

\noindent  
Finally we apply Lemma~\ref{lem:util-as-rev} with $h=4n$ and $\ell=1/2n$:
\begin{align}
\label{eq:integral} \log(8n^2) \expect[\alpha]{\alpha q(v)} & = u(v,\ell q)-u(v,hq) = u(v,p)-u(v,hq) + \frac 12 p(v)
\end{align}

\noindent  
Putting together Equations~\eqref{eq:rh}, \eqref{eq:alt-pricing}, and \eqref{eq:integral}, we get that for any buyer valuation $v$:
\begin{align*}
  \log(8n^2)\expect[\alpha]{\rev_v(\alpha q)} + 4\expect[T,a]{\rev_v(\gt)} & \ge \frac 12 p(v).
\end{align*}

\noindent  
The theorem now follows by taking expectations over the valuation function $v$ and recalling that $\{\alpha q\}$ and $\{\gt\}$ are all additive pricing functions.

\end{proof}

\section{Lower Bound}
\label{sec:lowerbound}

In this section we present our main lower bound, namely that the $O(\log n)$ approximation achieved in the previous section is tight in a very strong sense. We show that there exists a distribution over additive buyer types such that no mechanism with sub-exponential description complexity can $o(\log n)$-approximate the revenue from optimal deterministic Sybil-proof pricing. This implies in particular that no simple mechanism can achieve better than logarithmic approximation, including popular mechanisms studied in previous work such as selling separately, selling the grand-bundle, partition mechanisms or any combination of these mechanisms.

Before presenting our results for additive buyers, we first consider the simpler case where buyers are single-minded, i.e. they are only interested in purchasing a single set.

Our construction proceeds by identifying a large class of subadditive functions that assign independently arbitrary values in $[n^{1/6},n^{1/3}]$ to exponentially many sets $S_1,S_2,\cdots,S_N$. For any function $p$ in this class, pricing according to $p$ extracts full surplus from a distribution over single minded buyers where a buyer wanting set $S_i$ has value $p(S_i)$. Getting an $O(\log n)$ approximation for such a distribution as in the previous section is easy by setting the same price for every subset since there are only $O(\log n)$ different scales of prices to choose from. However, to obtain a better approximation one would need to charge high or low prices at different sets which requires at least one bit per set to describe. As there are exponentially many sets, mechanisms with subexponential description complexity cannot obtain better than logarithmic approximation.

%When facing a buyer whose value for those 
%Briest et al.~\cite{BCKW-SODA10} provided a matching lower bound on \sybilrev/\drev\ (equivalently \sybilrev/\srev\ since they studied unit-demand buyer). It is worth noting that our result gives a lower bound on the gap of \subaddrev\ and any mechanism with sub-exponential description complexity (including some popular benchmarks in previous literature, for example \srev, \brev, $\max(\srev,\brev)$), and not just \sybilrev/\srev. 

%The main idea behind this lower bound is that there exist (additive) valuation functions and subadditive pricings that extract almost the entire value, that are parameterized by exponentially many independent parameters. To be more specific, we find exponential number of sets $S_1,S_2,\cdots,S_N$ such that a function $p$ partially defined on theses sets with $p(S_i)$ being arbitrary value in a range of $\poly(n)$ can always be extended to a monotone subadditive pricing function. Previous section shows obtaining $O(\log n)$ approximation in revenue for each set is possible since we can guess $O(\log n)$ scales of prices.  Obtaining a better approximation would essentially amount to achieving a good quality compression of these parameters into sub-exponentially many bits, an impossible task. 

\subsection{The basic construction and a lower bound for single-minded buyers}

Let $p$ and $q$ be any two functions defined over the subsets of $[n]$. Let $\setdist$ be a demand distribution over sets, $\setdist\in\Delta(2^{[n]})$. We will say that $q$ {\em $c$-approximates $p$ from below over $\setdist$} if the following holds:
\[
\expect[S\sim\setdist]{q(S) \mathbbm{1}_{q(S)\le p(S)}} \ge \frac 1c \expect[S\sim\setdist]{p(S)}.
\]
Our argument will proceed in two parts. First, we will show that for any ``small'' class of functions $q$, there exists a subadditive (in fact, submodular) $p$ and a distribution $\setdist$ such that no $q$ in the class can $o(\log n)$-approximate $p$ from below over $\setdist$. Then we will show that for any $p$ and $\setdist$ of the form constructed in the first step, there exists a distribution over valuation functions such that the optimal subadditive revenue is a constant fraction of $\expect[S\sim\setdist]{p(S)}$, while the revenue of any other function $q$ is exactly $\expect[S\sim\setdist]{q(S) \mathbbm{1}_{q(S)\le p(S)}}$ for single-minded buyers. Together this will imply the following theorem.

\begin{theorem}
\label{thm:lb-singleminded}
  For any large enough $n$, there exists a distribution $\dist$ over single-minded valuation functions over $n$ items and a deterministic submodular pricing function $p$ such that $\rev_\dist(p)$ is a factor of $\Omega(\log n)$ larger than the revenue of any pricing that can be described using $2^{o(n^{1/6})}$ bits.
\end{theorem}

Let us begin by describing the class of subadditive functions $p$ that we will use in our argument. Let $\mathcal S = \{S_1, S_2, \cdots, S_N\}$ be a collection of $N$ subsets of $[n]$. Let $\beta = (b_1, \cdots, b_N)$ be a vector of integers of size $N$, where each coordinate is picked from the range $[\bmin,\bmax]$. We define a partial function $\fp$ as follows: for all $S_i\in \mathcal S$, set $\fp(S_i)=b_i$. The following lemma follows from the work of Balcan and Harvey~\cite{balcan2011learning} and shows that we can pick both $N$ and $\bmax/\bmin$ to be sufficiently large while ensuring that $p$ is submodular.\footnote{Matroid rank functions are a subclass of monotone submodular functions, which in turn are a subclass of all monotone subadditive functions.} See the appendix for a proof.

\begin{lemma}\label{thm_matroid}
Let $N=2^{n^{1/6}/8}$, $\bmin=n^{1/6}$, and $\bmax=n^{1/3}$. Then, there exists a collection of sets $\mathcal S = \{S_1, S_2, \cdots, S_N\}$, such that for each $i$ $|S_i|=n^{1/3}$; for each $i\neq j$, $|S_i\cap S_j|\leq n^{1/6}$; and for any integral vector $\beta\in [\bmin,\bmax]^N$, the partial function $\fp$ can be completed to a matroid rank function.
\end{lemma}

Let $\setdist$ be the uniform distribution over the collection $\mathcal S$. Our next lemma argues that for any small class of functions $\mathcal Q$, there exists a vector $\beta$, such that no function in $\mathcal Q$ can $c$-approximate $\fp$ from below over $\setdist$. Observe that because $\setdist$ only places non-zero mass over sets in $\mathcal S$, we do not need to specify a completion of $\fp$ for this lemma.

\begin{lemma}\label{thm-lbmain}
Let $m=\log(\bmax/\bmin)$, $c\le m/8$, and $\mathcal Q$ be an arbitrary class of functions defined over the subsets of $[n]$ with $|\mathcal Q|\le 2^{o(N/4^m)}$. Then there exists an integral vector $\beta\in [\bmin,\bmax]^N$ such that no function $q\in \mathcal Q$ can $c$-approximate $\fp$ from below over $\setdist$.
%If the value of $p(S_i)$ can be set to arbitrary value in $[\bmin,2^m\bmin]$.
%There exists $p$ such that for any $c=o(m)$, there exists no function $q$ that can be described by $2^{o(N4^{-m})}$ bits can $c$-approximate function $p$.
\end{lemma}
\begin{proof}
  Fix a function $q\in \mathcal Q$. We will pick $\beta$ from a distribution and show that the probability that $q$ $c$-approximates the corresponding function $\fp$ is small. For each $i\in [N]$, draw $b_i$ independently according to the following truncated geometric distribution: $\Pr[b_i=2^{k}\bmin]=\frac{2^{-k}}{1-2^{-m}}$ for $1\leq k\leq m$. Let $h_i = q(S_i)\mathbbm{1}_{q(S_i)\leq \fp(S_i)}$ be a random variable that depends on $\beta$. Then the statement that $q$ $c$-approximates $\fp$ from below over $\setdist$ is equivalent to the statement that 
\begin{align}
\label{eq:approx} \sum_{i\in [N]} h_i \ge \frac 1c \sum_{i\in [N]} \fp(S_i).
\end{align}

Observe that over the randomness in $\beta$, the variables $h_i$ are independent and bounded. For all $i$, $h_i\le\fp(S_i)\le 2^m\bmin$. Furthermore,  
\[\expect{h_i}=q(S_i)\Pr[\fp(S_i)\geq q(S_i)]\leq q(S_i) \frac{2\bmin}{q(S_i)} = 2\bmin.\]
On the other hand,
\[\expect{\fp(S_i)}=\expect{b_i}=\frac{m}{1-2^{-m}}\bmin.\] 
We can now bound the probability that \eqref{eq:approx} holds by applying concentration to the sums of $h_i$ and $\fp(S_i)$ respectively.
% (NEEDS TO BE REWORKED.)
% Consider the following random instance of valuation. First for each $i\in [N]$, randomly draw $t_i$ according to the following truncated geometric distribution: $\Pr[t_i=2^{k}\bmin]=\frac{2^{-k}}{1-2^{-m}}$ for $1\leq k\leq m$. Let $p(S_i)=t_i$. Notice that $q$ can $c$-approximate $p$ if and only if $\sum_{i=1}^{N}p(S_i)\leq\sum_{i=1}^{N}q(S_i)\mathbbm{1}_{q(S_i)\leq p(S_i)}$. Define $h_i=q(S_i)\mathbbm{1}_{q(S_i)\leq p(S_i)}$. Over the randomness of $p$, $h_1$, $h_2$, $\cdots$, $h_N$ are independent random variables. Observe that for every $i$, if $2^{k-1}\bmin<q(S_i)\leq 2^k\bmin$, then 
% \[\expect{h_i}=q(S_i)\Pr[p(S_i)\geq q(S_i)]\leq q(S_i)\Pr[p(S_i)\geq 2^k\bmin]\leq 2^k\bmin\cdot 2^{-k+1}=2\bmin,\]
% where the last inequality comes from $\Pr[p(S_i)\geq 2^k\bmin]\leq 2^{-k+1}$, thus $\expect{h_i}$ is always upper bounded by $2\bmin$. However, since 
% \[\expect{p(S_i)}=\expect{t_i}=\frac{m}{1-2^{-m}}\bmin,\] 
% the expected gap of $p$ and $q$ is $\Omega(m)$. Thus the probability that $q$ can $c$-approximate $p$ is
\begin{eqnarray*}
\Pr_\beta\left[\sum_i h_i \right. & \geq& \left. \frac{1}{c}\sum_i \fp(S_i)\right]\\
&\leq&\Pr\left[\sum_i h_i\geq\frac{1}{2c}\expect{\sum_i \fp(S_i)}\right]+\Pr\left[\sum_i \fp(S_i)<\frac{1}{2}\expect{\sum_i \fp(S_i)}\right]\\
&=&\Pr\left[\sum_i h_i\geq\frac{m}{2c(1-2^{-m})}N\bmin\right]+\Pr\left[\sum_i \fp(S_i)<\frac{m}{2(1-2^{-m})}N\bmin\right]\\
&\leq&\exp\left(-\frac{2(\frac{m}{2c(1-2^{-m})}N\bmin-2N\bmin)^2}{N(2^m\bmin)^2}\right)+\exp\left(-\frac{2(\frac{m}{2(1-2^{-m})}N\bmin)^2}{N(2^m\bmin)^2}\right)\\
&=&%O(2^{-N4^{-m}m^2/c^2})<
O(2^{-N4^{-m}}).
\end{eqnarray*}
Here the third line uses Hoeffding's inequality by observing that $0\leq h_i,\fp(S_i)\leq 2^m\bmin$; the last line follows using $c\le m/8$. The lemma now follows by taking the union bound over all $q\in \mathcal Q$.
% Thus the probability that a function $q$ described by $o(N4^{-m})$ bits can get revenue within factor $c=o(\log n)$ of $\subaddrev$ is at most $2^{o(N4^{-m})}O(2^{-N4^{-m}})<1$, therefore there exists a realization of $p$ where $p$ cannot be $c$-approximated by $q$ with sub-exponential description complexity.
\end{proof}

\noindent
We are now ready to prove Theorem~\ref{thm:lb-singleminded}.

\begin{proofof}{Theorem~\ref{thm:lb-singleminded}}
  Let $\bmin$, $\bmax$, $N$, and $\mathcal S$ be as given in Lemma~\ref{thm_matroid}. Let $\setdist$ be the uniform distribution over $\mathcal S$. Let $m=\log(\bmax/\bmin)$, $c=m/8=\Omega(\log n)$, and observe that the class $\mathcal Q$ of all pricings that can be described using $2^{o(n^{1/6})}$ bits has size $2^{o(N/4^m)}$. So we can apply Lemma~\ref{thm-lbmain} to obtain a vector $\beta$.

Now we will define a distribution $\dist$ over valuation functions as follows. For $i\in [N]$, let $v_i$ be the function that takes on value $b_i$ over any superset of $S_i$ and value $0$ otherwise. Function $v_i$ is instantiated with probability $1/N$. Let $p$ be the completion of $\fp$ as given by Lemma~\ref{thm_matroid}. Consider a pricing function that assigns a price of $\frac 12 p(S)$ to any set $S\subseteq [n]$. Since this function is subadditive and a buyer with value $v_i$ is single-minded and can afford to buy the set $S_i$, the mechanism obtains a revenue of $\frac 12 p(S_i)$ from this buyer. The mechanism's expected revenue over the distribution $\dist$ is then $\frac 12 \expect[S\sim\setdist]{\fp(S)}$.

On the other hand, for any monotone pricing function $q$, a buyer with value function $v_i$ purchases the set $S_i$ if and only if $q(S_i)\le v_i(S_i)=b_i=\fp(S_i)$. Therefore, the revenue of such a function $q$ over $\dist$ is at most
$\expect[S\sim\setdist]{q(S) \mathbbm{1}_{q(S)\le p(S)}}$.

The theorem now follows by applying Lemma~\ref{thm-lbmain}.
\end{proofof}

\subsection{A lower bound for additive buyers}
We now extend our lower bound to additive value buyers. Our construction is similar to that for Theorem~\ref{thm:lb-singleminded} but there are some subtle differences. As before, let $\mathcal S = \{S_1, S_2, \cdots, S_N\}$ and $\beta = (b_1, \cdots, b_N)\in [\bmin,\bmax]^N$. 
% Define $\fp$ over the sets in $\mathcal S$ as: $\fp(S_i)=b_i$ for all $S_i\in \mathcal S$. Let us extend this function to arbitrary sets in the natural way: $\fp(S) = \min_{A\subseteq [n]: \cup_{i\in A}S_i\superseteq S} \sum_{i\in A}\fp(S_i)$.
%We require a slightly different construction of the family $\mathcal S$ for additive buyers. The following lemma is proved in the appendix.
%\begin{lemma}\label{lem-set}
%There exists a collection of $N=\frac{1}{2}\cdot2^{n^{1/4}/8}$ sets, $\mathcal S=\{S_1,\cdots,S_{N}\}$, with $|S_i|=n^{1/2}$ for all $i\in [N]$ such that $\forall 1\leq i<j\leq N$, $|S_i\cap S_j|\leq n^{1/4}$. %Furthermore, let $\bmin=n^{1/4}$ and $\bmax=n^{1/2}$. Then, for any vector $\beta\in [\bmin,\bmax]^N$, the function $\fp$ is subadditive.
%\end{lemma}
We now state and prove our lower bound for additive buyers.
\begin{theorem}
\label{thm:lb-additive}
  For any large enough $n$, there exists a distribution $\dist$ over additive valuation functions over $n$ items and a deterministic monotone subadditive pricing function $p$ such that $\rev_\dist(p)$ is a factor of $\Omega(\log n)$ larger than the revenue of any mechanism that can be described using $2^{o(n^{1/6})}$ bits.
\end{theorem}
\begin{proof}
  Let $N$ and $\mathcal S$ be as defined in Lemma~\ref{thm_matroid}. Let $\bmin=2n^{1/6}$, $\bmax=n^{1/3}$, and fix an integral vector $\beta\in [\bmin,\bmax]^N$. Consider the following distribution $\dist_\beta$ over value functions. For $i\in [N]$, $v_{i,\beta}$ is an additive function that takes on the value $b_i/|S_i|$ over all items in $S_i$, and $0$ on items not in $S_i$. Observe that $v_{i,\beta}$ is a uniform additive valuation and $v_{i,\beta}(S_i) = b_i$. Function $v_{i,\beta}$ is instantiated with probability $1/N$. 

Now consider the pricing function $\fp$ defined over the sets in $\mathcal S$ as $\fp(S_i)=\frac 12 b_i$ for all $S_i\in \mathcal S$, and extended to arbitrary sets in the natural way: $\fp(S) = \min_{A\subseteq [N]: \cup_{i\in A}S_i\supseteq S} \sum_{i\in A}\fp(S_i)$. Observe that $\fp(S_i)<v_{i,\beta}(S_i)$. We claim that $\fp$ is subadditive and therefore extracts revenue $\frac 12 b_i$ from the buyer with type $v_{i,\beta}$. To see this, recall that the buyer obtains positive utility from the set $S_i$. Suppose the buyer instead decides to buy the collection $S_{i_1},S_{i_2},\cdots,S_{i_k}$. Since $|S_{i_j}\cap S_i|\leq n^{1/6}$, $v_{i,\beta}(S_{i_1}\cup S_{i_2}\cup\cdots\cup S_{i_k})\leq kn^{-1/6}b_i$. Thus his utility from buying these sets is at most 
\[kn^{-1/6}b_i-\sum_{j=1}^{k}\frac{1}{2}b_{i_j}\leq kn^{-1/6}\bmax-\frac 12 k \bmin =0.\] 
Therefore, we have $\fp(v_{i,\beta}) = \fp(S_i)$.

Now let $\mathcal Q$ be the class of mechanisms/pricings in the statement of the theorem and fix any $q\in\mathcal Q$. We will again define a distribution over instances by defining a distribution over the vectors $\beta$, as in the proof of Lemma~\ref{thm-lbmain}. Let $m=\log(\bmax/\bmin)$. For all $i\in [N]$, draw $b_i$ independently from the following truncated geometric distribution: $\Pr[b_i=2^{k}\bmin]=\frac{2^{-k}}{1-2^{-m}}$ for $1\leq k\leq m$.

Let $h_i = q(v_{i,\beta})$ be the revenue $q$ obtained from the buyer with value function $v_{i,\beta}$. Observe that $h_i$ is a random variable that depends on $b_i$. As before, the variables $h_i$ are independent and bounded by $2^m\bmin$. Furthermore, we can bound the expectation of $h_i$ by observing that a buyer with valuation $v_{i,\beta}$ is an additive buyer with uniform values over items in $S_i$. Selling a subset of items to this buyer is equivalent to selling fractional amounts of a single item to a single-parameter buyer. The optimal revenue from this buyer, over the randomness in $b_i$, is bounded by the revenue of single posted price. Therefore, $\expect{h_i}\le 2\bmin$. We now apply the same concentration argument as in the proof of Lemma~\ref{thm-lbmain} to obtain
\[\Pr_{\beta}\left[\sum_i h_i\geq\frac{1}{c} \sum_i \fp(S_i)\right] \le O(2^{-N4^{-m}}).\]
The theorem now follows by taking the union bound over $q\in \mathcal Q$.
\end{proof}

\bibliographystyle{plainnat}
\bibliography{newref,welfare,scheduling,agt,bmd,ea}

%\newpage
%\appendix
\appendix
\section{Deferred proofs}
\label{sec:deferred}

\subsection{Proof of Lemma~\ref{thm_matroid}}
\begin{numberedlemma}{\ref{thm_matroid}}
Let $N=2^{n^{1/6}/8}$, $\bmin=n^{1/6}$, and $\bmax=n^{1/3}$. Then, there exists a collection of sets $\mathcal S = \{S_1, S_2, \cdots, S_N\}$, such that for each $i$ $|S_i|=n^{1/3}$; for each $i\neq j$, $|S_i\cap S_j|\leq n^{1/6}$; and for any integral vector $\beta\in [\bmin,\bmax]^N$, the partial function $\fp$ can be completed to a matroid rank function.
\end{numberedlemma}
\begin{proof}
We need the following lemmas from \cite{balcan2011learning}.
\begin{lemma}\label{thm_bh}
(Theorem 9 of \cite{balcan2011learning}) Let $\mu$ and $\tau$ be non-negative integers. $f:2^{[N]}\to\mathbb{R}$ is called $(\mu,\tau)$-large if $f(J)\geq 0$, $\forall |J|<\tau$; $f(J)\geq\mu$, $\forall \tau\leq |J|\leq 2\tau-2$. Then for any sets $S_1,\cdots,S_N\subseteq[n]$, $\mathcal{I}=\{I:|I|\leq \mu\wedge|I\cap S(J)|\leq f(J),\forall J\subseteq[N],|J|<\tau\}$ is the family of independent sets of a matroid, here $S(J)=\bigcup_{j\in J}S_j$.
\end{lemma}

\begin{lemma}
(Theorem 13 of \cite{balcan2011learning}) Let $G(U\cup V,E)$ be bipartite graph. $G$ is called a $d,L,\epsilon$-lossless expander if $\Gamma({u})=d$, $\forall u\in U$; $\Gamma(J)\geq (1-\epsilon)d|J|$, $\forall J\subseteq U, |J|\leq L$. Then if $|U|=N$, $|V|=n$, $d\geq 2\log N/\epsilon$, $n\geq 6Ld/\epsilon$, a $(d,L,\epsilon)$-lossless expander exists.
\end{lemma}

We prove that for $N=2^{n^{1/6}/8}$, $d=b_{\max}=\mu=n^{1/3}$, $b_{\min}=n^{1/6}$, $\epsilon=\frac{b_{\min}}{4\mu}=\frac{1}{4n^{1/6}}$, $\tau=\frac{2\mu}{b_{\min}}=2n^{1/6}$, $L=2\tau=4n^{1/6}$, there exists $S_1,\cdots,S_N$ such that for any $b_1,\cdots,b_N\in[b_{\min},b_{\max}]$, $f(S_i)=b_i$ is a matroid rank function.

Let $S_j=\Gamma(\{j\})$ in the lossless expander (can check feasible under parameters above). Then $|S_j|=d=n^{1/3}$, $|S_i\cap S_j|=|S_i|+|S_j|-|\Gamma(\{i,j\})|\leq 2d-(1-\epsilon)2d<n^{1/6}$. Let $h(J)=\sum_{j\in J}b_j-\left(\mu|J|-|S(J)|\right)$.

\begin{lemma}
$h$ is $(\mu,\tau)$-large.
\end{lemma}
\begin{proof}
Notice that $h(J)\geq b_{\min}|J|-\epsilon|J|\mu=3\epsilon\mu|J|$. When $|J|\leq \tau$, $h(J)\geq 0$; when $|J|>\tau$, $h(J)\geq3\epsilon\mu\tau=1.5\mu>\mu$.
\end{proof}

By Theorem \ref{thm_bh}, $\mathcal{I}=\{|I|\leq \mu \wedge |I\cap S(J)|\leq h(J), \forall |J|< \tau\}$ is the family of independent sets of a matroid. For any $S_i$, pick $B_i\subseteq S_i$ such that $|B_i|=b_i$. Now we show that $B_i$ is a maximum independent subset of $S_i$. We need to verify the following properties:
\begin{itemize}
	\item $rank(S_i)\leq b_i$. This is true since for any $S\subseteq S_i$, if $S$ is independent, then $|S\cap S_i|\leq h(\{i\})=b_i$.
	\item $B_i$ is independent. Only need to show that for any $J$, $|B_i\cap S(J)|\leq h(J)$. When $i\in J$, $|J|\geq 2$, $h(J)=b_i+\sum_{j\neq i,j\in J}b_j-\epsilon\mu|J|\geq b_i+(|J-1|)b_{\min}-|J|\epsilon\mu\geq b_i$ since $\epsilon\mu=\frac{1}{4}b_{\min}$. When $i\not\in J$, $|B_i\cap S(J)|\leq\sum_{j\in J}|B_i\cap S_j|\leq \sum_{j\in J}|S_i\cap S_j|\leq |J|\cdot2\epsilon\mu<3\epsilon\mu|J|\leq h(J)$. Thus $B_i\in\mathcal{I}$, then $B_i$ is independent.
\end{itemize}

Thus for any $\mathbf{b}\in[b_{\min},b_{\max}]^N$, $f(S_i)=b_i$ can be extended to a feasible matroid rank function.
\end{proof}

\section{A constant upper bound for special demand distributions}
\label{sec:better}

In Section \ref{sec:lowerbound} we proved that we cannot obtain an $o(\log n)$-approximation to deterministic subadditive pricing via simple mechanisms for arbitrary distributions over valuation functions. However, when the distributions satisfy a certain property simple mechanisms, in particular item or bundle pricings, are able to do better.

\begin{definition}
  Given a deterministic Sybil-proof pricing $p$ and a distribution $\dist$ over valuations, the {\em demand distribution} $\Pi_{p,\dist}$ is the distribution over sets of items that specifies the random set of items bought by a buyer with value drawn from $\dist$ under pricing $p$: for all $S\subseteq [n]$, $\Pi_{p,\dist}(S) = \Pr_{v\sim\dist}[S = \argmax_{T\subseteq [n]} (v(T)-p(T))]$.\footnote{Generally speaking, specifying the demand distribution requires specifying a tie-breaking rule between multiple sets of equal utility. We will focus here on single-minded buyers, so the possibility of tie-breaking will not arise.}
\end{definition}

Observe that for any pricing $p$ and value distribution $\dist$, the revenue of the pricing is precisely $\expect[v\sim\dist]{p(v)} = \expect[S\sim\Pi_{p,\dist}]{p(S)}$. We will now show that if the buyer is single-minded and the demand distribution is a product distribution over items, then the revenue of $p$ can be approximated by a simple pricing.

% Let $f:\{0,1\}^n\to\R_0^+$ be a deterministic subadditive pricing function over sets. Assume that given such pricing, the set of items purchased by the buyer follows product distribution $\Pi$, where each item $i$ is purchased independently with probability $\pi_i$. While the subadditive function can be complicated, we prove that simple mechanism can get comparable revenue. To be more precise, either item pricing or grand bundle pricing is constant approximation of the revenue of pricing $f$.
\begin{theorem}
Given a deterministic monotone subadditive pricing function $p$ and value distribution $\dist$ over single-minded values, suppose that the demand distribution $\Pi_{p,\dist}$ is a product distribution over items, then $\rev_\dist(p)< 22.67\max(\srev_\dist,\brev_\dist)$.
\end{theorem}
\begin{proof}
As in the work of \cite{rw-15}, we will break up set of items and correspondingly the revenue obtained by $p$ into two components: over one of these components, a.k.a. the core, the price will concentrate around its expectation and can be approximated using a bundle pricing; over the other, a.k.a. the tail, a significant fraction of the revenue will be contributed by singleton items and can be recovered using an item pricing.

We write $\Pi$ as shorthand for $\Pi_{p,\dist}$. Number the items in decreasing order of their individual prices: $p(\{1\})\geq p(\{2\})\geq\cdots\geq p(\{n\})$. For items $i\in [n]$ let $\pi_i$ denote the marginal probability that $i$ is purchased, that is, $\pi_i = \Pr_{S\sim\Pi}[i\in S]$. Find the index $k\in[n]$ such that $\pi_1+\pi_2+\cdots+\pi_{k}<\frac{1}{2}$ and $\pi_1+\pi_2+\cdots+\pi_{k+1}\geq\frac{1}{2}$. The high value, low probability items $\{1, 2, \cdots, k\}$ will form the tail, and the remaining items will form the core.
%Based on the value of $\pi_{k+1}$, we have the following two cases. 
For any subset $S\subseteq [n]$, define $S_{\tail}=S\cap\{1,2,\cdots,k\}$ and $S_{\core}=S\cap\{k+1,\cdots,n\}$. By the subadditivity of $p$ we have $p(S)\leq p(S_{\tail})+p(S_{\core})$. Therefore, $\expect[S\sim\Pi]{p(S_{\tail})}+\expect[S\sim\Pi]{p(S_{\core})}$ is an upper bound of the revenue of $p$.

We will first bound the contribution of the core, $\expect[S\sim\Pi]{p(S_{\core})}$ using the following concentration lemma for subadditive functions from \cite{schechtman2003concentration}. To apply the lemma, we observe that $p(S_{\core})$ is Lipschitz with a Lipschitz constant of $\max_{i>k} p(\{i\}) = p(\{k+1\})$ because $S_{\core}$ contains only items with index larger than $k$.  
% Partition set $S$ purchased by the buyer into two parts: $S_{\tail}=S\cap\{1,2,\cdots,k+1\}$, and $S_{\core}=S\cap\{k+2,\cdots,n\}$. Since the pricing function $p$ is subadditive, for any set $S\subseteq [n]$, $f(S)\leq f(S_{\tail})+f(S_{\core})$. Thus $\expect{f(S_{\tail})}+\expect{f(S_{\core})}$ is an upper bound of the revenue of subadditive pricing. Let $c=f(\{k+1\})$. Since all items are sorted in decreasing order of the value of singleton sets, we observe that $f(S_{\core})$ is $c$-Lipschitz, since $S_{\core}$ contains only items with index larger than $k+2$. Similar to the approach in \cite{rw-15}, we apply the following concentration result on subadditive functions from \cite{schechtman2003concentration}.
\begin{theorem}
(Corollary 12 from \cite{schechtman2003concentration}) Suppose that $f(X)$ is a non-negative $c$-Lipschitz subadditive function, where $X\in\{0,1\}^n$ is drawn from a product distribution $D$. If $a$ is the median of $f$, then for any $k>0$, $\Pr_{X\sim D}[f(X)\geq 3a+k]\leq 2^{2-k/c}$.
\end{theorem}
Setting $a = \operatorname{median}_{S\sim\Pi}(p(S_\core))$ and $c=p(\{k+1\})$, we obtain
%Let $a$ be the median of $f(S_\core)$. Notice that
\[\expect[S\sim\Pi]{p(S_{\core})}=\int_{t\geq 0}\Pr[p(S_{\core})>t]dt\leq 3a+\int_{k\geq0}2^{2-k/c}dk=3a+\frac{4}{\ln 2}c.\]
We now observe that we can recover both of the terms above using a bundle pricing. In particular, by setting a constant bundle price of $a$, we obtain a revenue of  $a\Pr_{v\sim\dist} [v([n])>a] \ge a\Pr_{S\sim\Pi} [p(S)>a] \ge a\Pr_{S\sim\Pi} [p(S_\core)>a] = a/2$. On the other hand, by setting a constant bundle price of $c$, we obtain a revenue of $c\Pr_{v\sim\dist} [v([n])>c] \ge c\Pr_{S\sim\Pi} [p(S)>c] \ge c\Pr_{S\sim\Pi}[\exists i\le k+1: i\in S]$. The latter probability can be bounded from below as:
% Since $a$ is the median of $f(S_\core)$, consider a mechanism that sells the entire bundle of $n$ items at price $a$, then the revenue achieved by such mechanism is $\frac{1}{2}a$, which implies $\brev\geq\frac{1}{2}a$. Since $c=f(\{k+1\})$, then for every set $S$, if $S$ contains an item with index at most $k+1$, then $f(S)\geq c$. Consider a mechanism that sells the entire bundle of $n$ items at price $c$, then the probability of sell is at least
\[\Pr_{S\sim\Pi}[\exists i\leq k+1,\ i\in S]=1-(1-\pi_1)(1-\pi_2)\cdots(1-\pi_{k+1})\geq 1-\left(1-\frac{\sum_{i\leq k+1}\pi_i}{k+1}\right)^{k+1}\geq1-e^{-1/2},\]
where the first inequality follows by applying Jensen's inequality, and the second follows by using $\sum_{i\leq k+1}\pi_i\geq\frac{1}{2}$. 

%Thus the revenue collected from the bundling pricing mechanism is at least $(1-e^{-1/3})c$, which implies $\brev\geq (1-e^{-1/3})c$. 
Thus we can bound $\expect{p(S_\core)}$ as follows:
\[\expect{p(S_{\core})}\leq3a+\frac{4}{\ln 2}c\leq 6\,\brev+\frac{4}{(1-e^{-1/2})\ln 2}\brev< 20.67\,\brev.\]

Now we bound $\expect{p(S_\tail)}$. By subadditivity, 
\[\expect{p(S_\tail)}\leq\expect{\sum_{i\leq k}p(S_\tail\cap\{i\})}=\sum_{i\leq k}\pi_ip(\{i\}).\] 
Recall that $\pi_1+\pi_2+\cdots+\pi_{k}\leq\frac{1}{2}$. Consider the additive pricing that sells items in $\{1,2,\cdots,k\}$ at item prices $q_i=p(\{i\})$, and allocates items $i>k$ for free.

Consider a buyer that purchases a set $S$ under pricing $p$ with $S_\tail=\{i\}$. In the pricing $q$, the buyer continues to afford the set $S$ at a lower price of $q_i=p(\{i\})<p(S)$. Therefore, the revenue of the pricing $q$ from this buyer is at least $q_i=p(\{i\})$. Since this buyer is instantiated with probability at least $\Pr[S_\tail=\{i\}] \ge \frac 12 \pi_i$, we get that $\expect{p(S_\tail)}\leq 2\,\rev_\dist(q)\le 2\,\srev$. Then combining the above cases we get $\rev_\dist(p)\leq 22.67\max(\srev_\dist,\brev_\dist)$.
\end{proof}

\end{document}